\documentclass[preprintnumbers, prd, onecolumn, showpacs,floatfix,preprintnumbers,
superscriptaddress, nofootinbib]{revtex4}
\usepackage[papersize={8.5in,11in}]{geometry}
\usepackage{graphicx}
\usepackage{epsfig}
\usepackage{bm}
\usepackage{amssymb}
\usepackage{float}
\usepackage{amsmath}
\usepackage{dcolumn}
\usepackage{cancel}
\usepackage[colorlinks]{hyperref}
\usepackage[usenames,dvipsnames]{color}
\usepackage[spanish]{babel}
\hypersetup{
     breaklinks=true,
    pdfstartview={FitH},    
    colorlinks=true,       
    linkcolor=blue,          
    citecolor=red,        
    filecolor=magenta,      
    urlcolor=blue,           
    anchorcolor=green,      
    linktocpage=true
}

\begin{document}
\title{C{\'a}lculo de constantes {\'o}pticas de pel{\'i}culas delgadas de Silicio compensado a trav{\'e}s del m{\'e}todo de Swanepoel}

\author{W. A. Rojas C.}
\email{warojasc@unal.edu.co}
\affiliation{Department of Physics, Universidad Nacional de Colombia\\ sede Bogot{\'a}, postcode UN. 11001 \\ Colombia}

\author{M. Chaparro P.}
\email{mchaparrop@unal.edu.co}
\affiliation{Department of Physics, Universidad Nacional de Colombia\\ sede Bogot{\'a}, postcode UN. 11001 \\ Colombia}

\author{L. M. Pardo C.}
\email{lmpardoc@unal.edu.co}
\affiliation{Department of Physics, Universidad Nacional de Colombia\\ sede Bogot{\'a}, postcode UN. 11001 \\ Colombia}

\author{ M. I. Cruz F.}
\email{micruzf@unal.edu.co}
\affiliation{Department of Physics, Universidad Nacional de Colombia\\ sede Bogot{\'a}, postcode UN. 11001 \\ Colombia}

\begin{abstract}
Characterization and determination of the optical constants for thin film silicon compensated through the transmittance spectrum is presented. Such properties were determined by R. Swanepoel model. Comparison between the refractive indices of pure silicon and silicon compensated. Increasing refractive index behavior at low wavelengths in both plots was observed. The difference lies in the concentration and type of element that has been added to the  thin film. The behavior of the dielectric constant was studied, their relationship with the refractive index. It is found that the value of the dielectric constant to a thin film of silicon compensated agree with those reported in the literature. The reported value of the gap for pure silicon corresponds to $ 1.11eV $ and the value of the gap for the sample corresponds to $ 0.7275eV $, the discrepancy is due to the level of concentration in the compensated silicon film.

\end {abstract}

\maketitle
\section{Introducci{\'o}n}
Se presenta la caracterizaci{\'o}n y determinaci{\'o}n de las  constantes {\'o}pticas para una pel{\'i}cula delgada de Silicio compensado a trav{\'e}s del espectro de transmitancia. Tales propiedades fueron determinadas por el modelo de R. Swanepoel. El modelo propuesto por R. Swanepoel  proporciona una  forma para determinar las propiedades {\'o}pticas de las pel{\'i}culas delgadas, las propiedades estudiadas fueron el indice de refracci{\'o}n $n$, el espesor de la pel{\'i}cula $d$, el {\'i}ndice de absorci{\'o}n $\alpha$, el coeficiente de extinci{\'o}n $k$ y la brecha de energ{\'i}a prohibida {\'o}ptica (tambi{\'e}n conocida como gap) $E_{g}$ \cite{Swanepoel,Mesa}. Se contrastaron los valores del espesor $d$ y el experimental  $d_{p}$ que fue medido en un perfil{\'o}metro Dektak 150. A la pel{\'i}cula en estudio se le practicaron 5 ensayos de perfilometr{\'i}a en los que se  determin{\'o} que el espesor es $d_{p}=308,4998nm$ valor que fue comparado con el modelo $d=282,96nm\pm 8\%$.  La determinaci{\'o}n de los $T_{M}$ y $T_{m}$ se realiz{\'o} por interpolaci{\'o}n de Lagrange. Esto permiti{\'o} determinar una  ecuaci{\'o}n de  Cauchy de la  forma	
\begin{equation}
		n(\lambda)=\frac{596000}{\lambda^{2}}+3.4672.
		\label{Eq1}
\end{equation}
		De la comparaci{\'o}n entre los {\'i}ndices de refracci{\'o}n del Silicio puro y del Silicio compensado se  observ{\'o} el comportamiento creciente del {\'i}ndice de refracci{\'o}n para bajas longitudes de onda en ambas gr{\'a}ficas. La  diferencia  radica en la concentraci{\'o}n y los elementos que se le han añadido a la  pel{\'i}cula en estudio. Se estudio el comportamiento de la constante diel{\'e}ctrica, su relaci{\'o}n con el indice de refracci{\'o}n. Hall{\'a}ndose que el valor de la constante diel{\'e}ctrica para una pel{\'i}cula delgada de Silicio compensado esta de acuerdo con lo reportado en la literatura. 	El valor reportado del gap para el Silicio en puro corresponde a $1,11eV$ y el valor del gap para la muestra estudiada corresponde a $0,7275eV$, la discrepancia se debe al nivel del concentraci{\'o}n con que est{\'a}  compensada la pel{\'i}cula de Silicio.
\subsection{El modelo}
\begin{figure}[ht]
	\centering
			\includegraphics[width=0.5\textwidth]{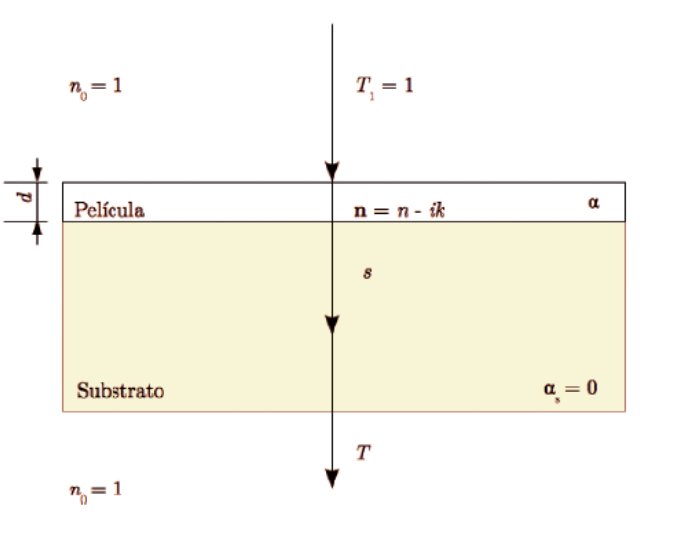}
		\caption{Sistema compuesto por una pel{\'i}cula delgada absorbente sobre un sustrato transparente finito \cite{Mesa}.}
	\label{F1}
\end{figure}

\begin{figure}[ht]
	\centering
			\includegraphics[width=0.5\textwidth]{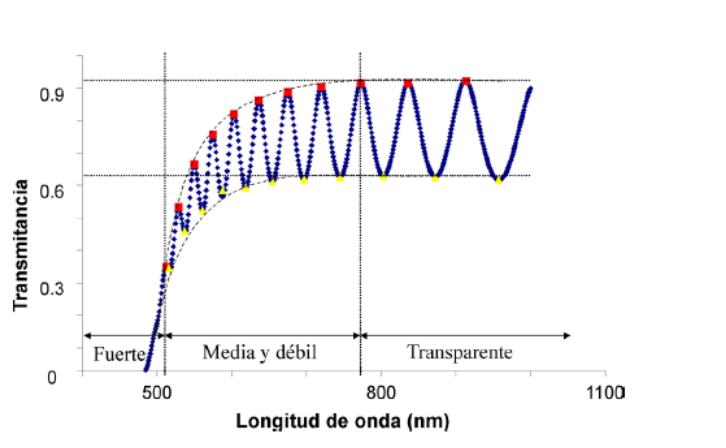}
		\caption{Espectro de transmitancia en el  cual se han definido tres  regiones de alta, media-baja y absorci{\'o}n nula, tal espectro se halla delimitado por dos curvas envolventes para los puntos de transmitancia m{\'a}ximos $T_{M}$ y m{\'i}nimos $T_{m}$ \cite{Mesa}.}
	\label{F2}
\end{figure}

En este punto se seguir{\'a} el an{\'a}lisis realizado por Mesa et al \cite{Mesa}. Swanepoel, presenta un procedimiento para calcular las constantes {\'o}pticas de pel{\'i}culas delgadas de materiales semiconductores amorfos a partir de datos obtenidos  experimentalmente del espectro de transmitancia. Este procedimiento asume un sistema compuesto por una pel{\'i}cula delgada homog{\'e}nea en espesor y con {\'i}ndice de refracci{\'o}n complejo $\eta= n - ik$   que es depositada sobre un sustrato transparente de {\'i}ndice de refracci{\'o}n $s$ y espesor mayor que el de la pel{\'i}cula, tal  como se puede ver en la Figura  \ref{F1}.
 La parte real $n$ del {\'i}ndice de refracci{\'o}n determina la velocidad con que la radiaci{\'o}n se propaga en el material y el factor $k$  es expresado en t{\'e}rminos  $\alpha$ de la forma
\begin{equation}
\alpha=\frac{4\pi k}{\lambda}.
\label{Eq2}
\end{equation}
Esto permite realizar un an{\'a}lisis de los efectos de interferencia que se observan en los espectros de transmitancia como consecuencia de la superposici{\'o}n de los haces reflejados y transmitidos en las interfaces pel{\'i}cula/aire y sustrato/pel{\'i}cula, obteniendo una expresi{\'o}n general para el valor de la transmitancia $T$ en funci{\'o}n de la longitud de onda $\lambda$ y de los par{\'a}metros $n$, $s$, $\alpha$ y espesor $d$
\begin{equation}
T=\frac{A}{B-Cx+Dx^2},
\label{Eq3}
\end{equation}
donde se tienen que los par{\'a}metros $A$, $B$, $C$ y $D$ son
\begin{equation}
A=16s\left[n^{2}+k^{2}\right],
\label{Eq4}
\end{equation}
\begin{equation}
B=\left\{ \left[n+1\right]^{2}+k\right\}\left\{ \left[n+1\right] \left[n+s^{2}\right]+k^{2}\right\},
\label{Eq5}
\end{equation}
\begin{equation}
	C=\left\{\left[n^{2}-1+k^{2}\right]\left[n^{2}-s^{2}+k^{2}\right]-2k^{2} \left[s^{2}+1\right]\right\}2\cos \phi
	-2k\left\{2\left[n^{2}-s^{2}+k^{2}\right]+\left[s^{2}+1\right]\left[n^{2}-1+k^{2}\right]\right\}\sin \phi,
	\label{Eq6}
\end{equation}
\begin{equation}
	D=\left\{\left[n-1\right]^{2}+k^{2}\right\}\left\{\left[n-1\right]\left[n-s^{2}\right]+k^{2}\right\},
	\label{Eq7}
\end{equation}
\begin{equation}
	\phi=\frac{4\pi d}{\lambda},
	\label{Eq8}
\end{equation}
\begin{equation}
	x=e^{-\alpha d}
	\label{Eq9}
\end{equation}
y
\begin{equation}
	\alpha=\frac{4\pi k}{\lambda}.
	\label{Eq10}
\end{equation}
El espectro puede ser dividido en cuatro regiones, tal como  se ve en la  Figura \ref{F2}. En la regi{\'o}n transparente $\alpha=0$ y la transmitancia se encuentra determinada por $n$ y $s$ a trav{\'e}s de las reflexiones m{\'u}ltiples. En la regi{\'o}n de absorci{\'o}n d{\'e}bil $\alpha$ es bajo y la transmitancia empieza a reducirse; en la regi{\'o}n de absorci{\'o}n media $\alpha$ crece y la transmitancia disminuye debido al efecto de $\alpha\neq 0$. En la regi{\'o}n de fuerte absorci{\'o}n, la transmitancia disminuye dr{\'a}sticamente debido, casi exclusivamente, a la influencia de $\alpha$.
Para una primera aproximaci{\'o}n, se consider{\'o} {\'u}nicamente el sustrato de modo que la transmitancia es considerada libre de interferencia de la forma
\begin{equation}
T_{s}=\frac{\left[1-R\right]^{2}}{1-R^{2}},
\label{Eq11}
\end{equation} 
donde $T_{s}$ corresponde  al coeficiente de transmitancia, incidiendo normalmente sobre la  superficie del sustrato y $R$ es el coeficiente de reflexi{\'o}n de la muestra o la  pel{\'i}cula
\begin{equation}
R=\left[\frac{s-1}{s+1}\right]^{2}.
\label{Eq12}
\end{equation}
Por lo que $T_{s}$ se reescribe como
\begin{equation}
T_{s}=\frac{2s}{s^2+1}.
\label{Eq13}
\end{equation}
Y el valor posible del {\'i}ndice de refracci{\'o}n $s$ para el sustrato queda
\begin{equation}
s=\frac{1}{T_{s}}\pm \sqrt{\frac{1}{T_{s}^{2}}-1},
\label{Eq14}
\end{equation}
en donde s{\'o}lo se debe  considerar la parte positiva de la anterior expresi{\'o}n. De otro lado, se tiene  que las  franjas de interferencia est{\'a}n descritas  de la  forma
\begin{equation}
2nd=m \lambda
\label{Eq15}
\end{equation}
donde $m=0,1,2\ldots$; que produce el patr{\'o}n de interferencia en pel{\'i}culas delgadas.  En tal sentido, Swanepoel propone una simplificaci{\'o}n en las expresiones, para un sistema compuesto por una pel{\'i}cula absorbente sobre un sustrato transparente finito y para el caso de un substrato asumido infinito. S{\'i} se considera que $k = 0$, que resulta ser una aproximaci{\'o}n v{\'a}lida en la mayor parte de la regi{\'o}n espectral \cite{Swanepoel,Mesa},  la transmisi{\'o}n $T$ est{\'a} dada por
\begin{equation}
T=\frac{Ax}{B-Cx \cos\varphi +Dx^{2}},
\label{Eq16}
\end{equation}
en donde los par{\'a}metros $A$, $B$, $C$, $D$, $\varphi$ y $x$ se reescriben como
\begin{equation}
A=16n^{2}s,
\label{Eq17}
\end{equation}
\begin{equation}
B=\left[n+1\right]^{3}\left[n+s^{2}\right],
\label{Eq18}
\end{equation}
\begin{equation}
C=2\left[n^{2}-1\right]\left[n^{2}+s^{2}\right],
\label{Eq19}
\end{equation}
\begin{equation}
D=2\left[n-1\right]^{3}\left[n+s^{2}\right],
\label{Eq20}
\end{equation}
\begin{equation}
\varphi=\frac{4\pi nd}{\lambda},
\label{Eq21}
\end{equation}
y 
\begin{equation}
x=e^{-\alpha d}.
\label{Eq22}
\end{equation}
La absorbancia $x$ para del sistema  t{\'e}rminos de la transmitancia libre de interferencia $T_{\alpha}$ de la  forma  
\begin{equation}
x=\frac{P+\sqrt{P^{2}+2QT_{\alpha}\left[1-R_{2}R_{3}\right]}}{Q},
\label{Eq23}
\end{equation}
en donde los par{\'a}metros $P$, $Q$, $R_{1}$, $R_{2}$ y $R_{3}$ son
\begin{equation}
Q=R_{1}R_{2}+R_{1}R_{3}-2R_{1}R_{2}R_{3},
\label{Eq24}
\end{equation}
\begin{equation}
P=\left[ R_{1}-1\right]\left[ R_{2}-1\right]\left[ R_{3}-1\right],
\label{Eq25}
\end{equation}
\begin{equation}
R_{1}=\left[\frac{1-n}{1+n}\right]^{2},
\label{Eq26}
\end{equation}
\begin{equation}
R_{2}=\left[\frac{n-s}{n+s}\right]^{2},
\label{Eq27}
\end{equation}
y
\begin{equation}
R_{3}=\left[\frac{s-1}{s+1}\right]^{2}.
\label{Eq28}
\end{equation}
Es posible  plantear dos funciones que describan  los extremos de las franjas de interferencia de la  forma
\begin{equation}
T_{M}=\frac{Ax}{B-Cx+Dx^{2}}
\label{Eq29}
\end{equation}
para los m{\'a}ximos $T_{M}$ de la Figura \ref{F2}. Y de la misma manera  para los  m{\'i}nimos $T_{m}$
\begin{equation}
T_{m}=\frac{Ax}{B+Cx+Dx^{2}}.
\label{Eq30}
\end{equation}
Este modelo ha repartido el espectro de transmitancia en tres  regiones de acuerdo al proceso de absorci{\'o}n en la pel{\'i}cula a saber:
\begin{enumerate}
		\item Regi{\'o}n transparente, definida  como aquella regi{\'o}n del espectro donde  se puede considerar que $\alpha=0,\,\,x=1$, por lo que la posici{\'o}n de los  $T_{M}$ queda definida como
	\begin{equation}
	T_{M}=\frac{2s}{s^{2}+1}.
	\label{Eq31}
	\end{equation}
	La posici{\'o}n de los m{\'i}nimos $T_{m}$ es
	\begin{equation}
	T_{m}=\frac{4n^{2}}{n^{4}+n^{2}\left(s^{2}+1\right)+s^{2}}.
	\label{Eq32}
	\end{equation}
	Y el {\'i}ndice de refracci{\'o}n de la pel{\'i}cula delgada  para esta regi{\'o}n es
	\begin{equation}
	n=\sqrt{M+\sqrt{M^{2}-s^{2}}},\,\,\,\,M=\frac{2s}{T_{m}}+\frac{s^{2}+1}{2}.
	\label{Eq33}
	\end{equation}
		\item Regi{\'o}n de baja y media absorci{\'o}n, para esta zona del espectro de transmitancia se tiene que $\alpha\neq 0, x<1$, por lo  que la relaci{\'o}n de la resta de sus inversos multiplicativos de \eqref{Eq29} y \eqref{Eq30}  es
		\begin{equation}
	\frac{1}{T_{m}}-\frac{1}{T_{M}}=\frac{2C}{A}.
	\label{Eq34}
	\end{equation}
		Y el {\'i}ndice de refracci{\'o}n de la pel{\'i}cula delgada  para esta regi{\'o}n es
			\begin{equation}
	n=\sqrt{N+\sqrt{N^{2}-s^{2}}},\,\,\,\,N=2s\left[\frac{T_{M}-T_{m}}{T_{M}T_{m}}\right]+\frac{s^{2}+1}{2}.
	\label{Eq35}
	\end{equation}
	la absorbancia $x$ en t{\'e}rminos de los $T_{M}$ es
	\begin{equation}
	x=\frac{E_{M}-\sqrt{E^{2}_{M}-\left(n^{2}-1\right)^{3}\left(n^{2}-s^{2}\right) }}{\left(n-1\right)^{3}\left(n-s^{2}\right)},\,\,\,\,E_{M}=\frac{8n^{2}s}{T_{M}}+\left(n^{2}-1\right)\left(n^{2}-s^{2}\right).
	\label{Eq36}
	\end{equation}
		La suma de los rec{\'i}procos de \eqref{Eq29} y \eqref{Eq30} conduce a
	\begin{equation}
	\frac{2T_{M}T_{m}}{T_{M}+T_{m}}=\frac{Ax}{B+Dx^{2}}.
	\label{Eq37}
	\end{equation}
	la absorbancia $x$ en t{\'e}rminos de los $T_{i}$ es
	\begin{equation}
	x=\frac{F-\sqrt{F^{2}-\left(n^{2}-1\right)^{3}\left(n^{2}-s^{2}\right) }}{\left(n-1\right)^{3}\left(n-s^{2}\right)},\,\,\,\,F=\frac{8n^{2}s}{T_{i}},\,\,\,\,T_{i}=\frac{2T_{M}T_{m}}{T_{M}+T_{m}},
	\end{equation}
	Donde $T_{i}$ representa una curva que pasa a trav{\'e}s de los puntos de inflexi{\'o}n de las franjas presentadas en la Figura \ref{F2}.  La transmitancia libre de interferencias $T_{\alpha}$ se calcula  como
	\begin{equation}
	T_{\alpha}=\frac{Ax}{\sqrt{\left[ B-Cx\cos\varphi +Dx^{2}\right]  \left[ B+Cx\cos\varphi +Dx^{2}\right]}}=\sqrt{T_{M}T_{m}}.
	\label{Eq39}
	\end{equation}
	donde se  tiene que $T_{\alpha}$ corresponde  a la media geom{\'e}trica de $T_{M}$ y $T_{m}$.
		\item Regi{\'o}n de intensa absorci{\'o}n, en esta situaci{\'o}n,  las franjas de interferencia desaparecen, por lo que no hay forma de calcular $n$ y $x$ independientemente en esta regi{\'o}n a partir del espectro de transmitancia {\'u}nicamente. Si se  desprecian los efectos  debido a la interferencia en  la pel{\'i}cula, bajo la  condici{\'o}n $x\ll 1$
	\begin{equation}
	T_{0}=\frac{Ax}{B},\,\,\,\,\,\,x\approx \frac{\left(n-1\right)^{3}\left(n+s^{2}\right)}{16n^{2}s}T_{0}.
	\label{Eq40}
	\end{equation}
	\end{enumerate}
\section{Montaje y equipos empleados}
\subsection{Espectro de transmitancia}
\begin{figure}[ht]
	\centering
			\includegraphics[width=0.5\textwidth]{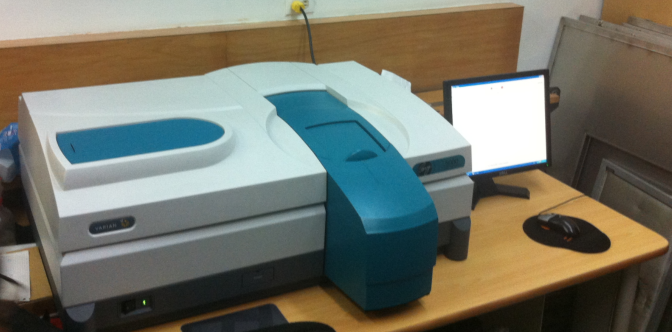}
		\caption{Espectrofot{\'o}metro Cary 5000 del Departamento de F{\'i}sica de la Universidad Nacional de Colombia.}
	\label{F3}
\end{figure}

\begin{figure}[ht]
	\centering
			\includegraphics[width=0.5\textwidth]{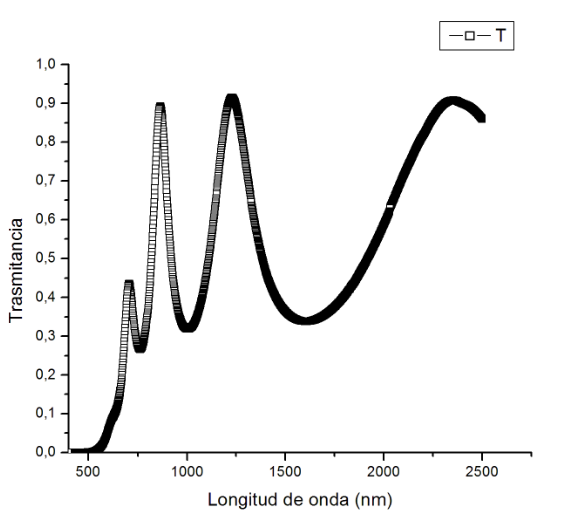}
		\caption{Espectro de transmitancia para una  pel{\'i}cula delgada de silicio compensado. Primera prueba.}
	\label{F4}
\end{figure}

\begin{figure}[ht]
	\centering
			\includegraphics[width=0.5\textwidth]{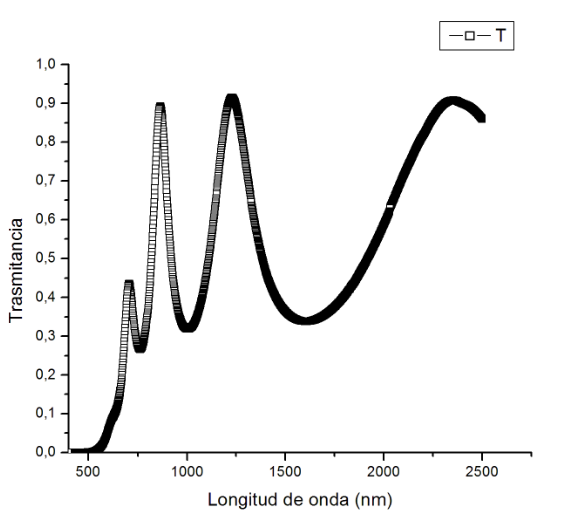}
		\caption{Segunda prueba.}
	\label{F5}
\end{figure}

\begin{figure}[ht]
	\centering
			\includegraphics[width=0.5\textwidth]{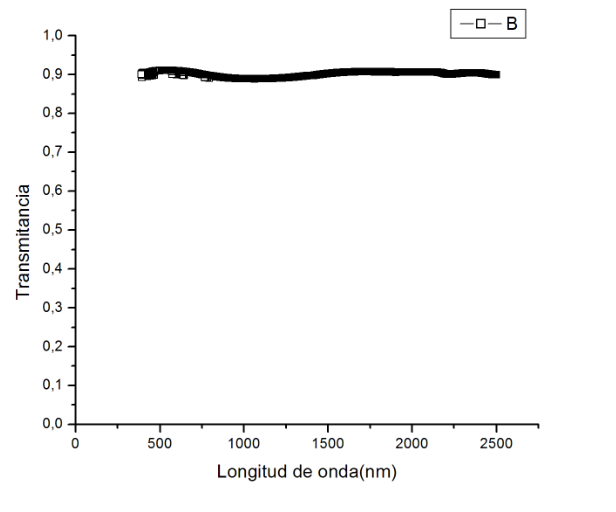}
		\caption{Espectro de transmitancia para el sustrato sin pel{\'i}cula delgada.}
	\label{F6}
\end{figure}

Para  determinar el tipo de espectro de transmitancia de la  muestra suministrada (Silicio compensado), se  emple{\'o} un espectrofot{\'o}metro UV-Vis-IR. El Cary 5000 es espectrofot{\'o}metro UV -Vis -NIR con rendimiento fotom{\'e}trico  el rango de $175 - 3300 nm$ y con una margen de error $\Delta \lambda=0,02nm$. El uso de un detector de PbSmart, el Cary 5000 ampl{\'i}a su gama NIR a 3300 nm y es controlado por el software Cary WinUV, que se halla disponible en las instalaciones del Departamento F{\'i}sica de la Universidad Nacional de Colombia, tal como se  halla en la  Figura \ref{F3}.
Con este  equipo y la muestra de  Silicio compensado se obtuvo el espectro de  transmitancia de la pel{\'i}cula que se puede ver en la  Figura \ref{F4}.
El proceso se repiti{\'o} dos veces para verificar el espectro de la pel{\'i}cula, el cual se puede ver en la Figura \ref{F5}.
De lo anterior se observa,  que los dos espectros tomados  a la  misma muestra son similares, se opta por caracterizar la pel{\'i}cula de muestra  con el espectro de la primera prueba, ver Figura \ref{F4}. Para determinar el {\'i}ndice de refracci{\'o}n del sustrato $s$, se realiz{\'o} un espectro de transmitancia al sustrato (vidrio) por separado, es decir sin pel{\'i}cula,  del  cual se obtuvo el  espectro que se observa en la Figura \ref{F6}.
\subsection{Espesor de la pel{\'i}cula delgada de Silicio compensado.}
\begin{figure}[ht]
	\centering
			\includegraphics[width=0.5\textwidth]{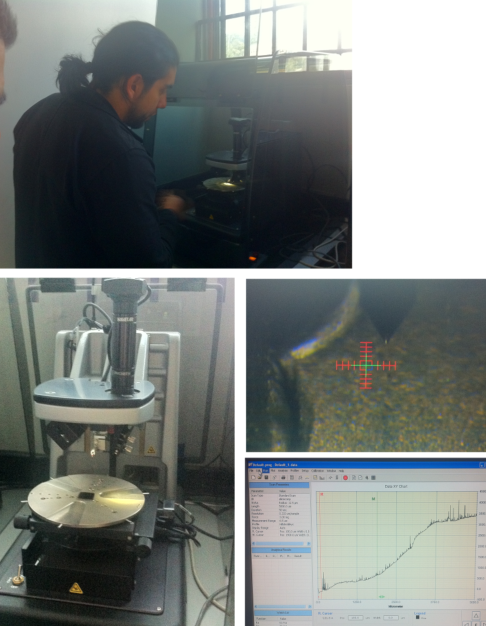}
		\caption{Equipo de perfilometr{\'i}a disponible en Departamento de F{\'i}sica de la Universidad Nacional de Colombia.}
	\label{F7}
\end{figure}

El perfil{\'o}metro es un equipo que puede tomar espesores y rugosidades de pel{\'i}culas delgadas, del orden de las micras, opera mediante una torre donde tiene una c{\'a}mara (zoom 166x),  con una punta (diamante) y un sensor de proximidad. La punta hace contacto con la muestra, mediante el brazo hace un barrido desde el sustrato hasta la muestra, a medida que la punta de diamante hace contacto con la superficie, en su brazo tiene un solenoide de fuerza, este sirve para ir graduando como debe interactuar la punta con la muestra.  El perfil{\'o}metro toma la señal de espesor de la pel{\'i}cula mediante un sensor (Low-Inertia Sensor LIS 3) que se denomina LVDT, y este opera haciendo que las variaciones mec{\'a}nicas se conviertan en señales de corriente continua que  son enviadas al software para ser interpretadas como espesores de la muestra. Despu{\'e}s de hacer el barrido y enviar las señales al software, este presenta una imagen del espesor de la muestra tomando como referencia el sustrato, de acuerdo con esto, se puede realizar por medio del software una aproximaci{\'o}n promedio del espesor de la pel{\'i}cula. De igual manera, presenta la distancia vertical punto a punto del sustrato y de la muestra,  ver Figura \ref{F7}.

\begin{figure}[ht]
	\centering
			\includegraphics[width=0.5\textwidth]{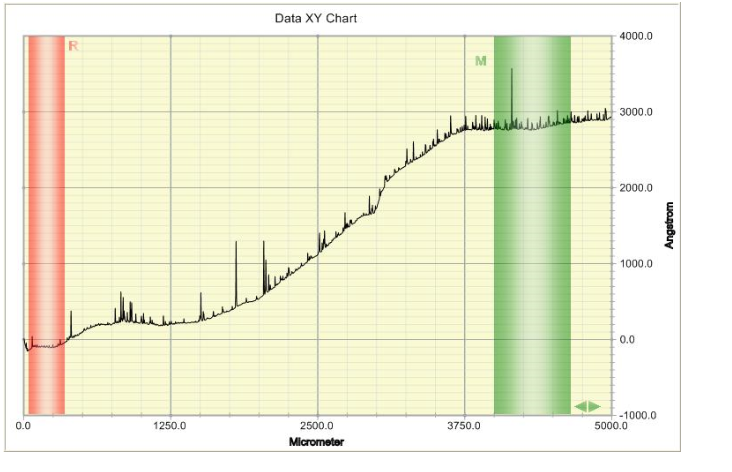}
		\caption{Curva t{\'i}pica para la  muestra en estudio.}
	\label{F8}
\end{figure}

A la pel{\'i}cula  en estudio se le hicieron 5 ensayos para determinar su espesor experimental $\left\langle d_{p}\right\rangle= 308.4998nm$. Una curva t{\'i}pica que muestra el equipo de perfilometr{\'i}a para la muestra en estudio se puede ver en la Figura \ref{F8}.

\section{An{\'a}lisis de Resultados}
Una  vez obtenidos el espesor de la  pel{\'i}cula y los espectros en el rango
\begin{equation}
	300 nm \leq \lambda \leq2500 nm, 
	\label{Eq41}
	\end{equation}
para tal intervalo, de  acuerdo  con el espectro de transmitancia del sustrato que se observa en la  Figura 6, se advierte que la transmitancia no fluct{\'u}a de manera considerable para el rango de longitud de onda establecido. Por lo que se  considera tal  transmitancia constante
\begin{equation}
\left\langle T_{s}\right\rangle=0.9003\pm  0.0066.
\label{Eq42}
\end{equation}
Lo  cual  permite obtener el {\'i}ndice de refracci{\'o}n del sustrato  a partir de \eqref{Eq13}
\begin{equation}
s=1.5142.
\label{Eq43}
\end{equation}
Del espectro que se aprecia en la Figura \ref{F4}. se obtuvieron los siguientes  valores de $T_{M}$ y $T_{m}$
\begin{table}[!hbt]
\begin{center}
\begin{tabular}{|l | l | l|}
\hline
$\lambda(nm)$	 	& $T_{M}$		&$T_{m}$ \\   
\hline 
		707.0115		& 		0.4355	&	$-$		\\   
\hline 
			866.0608	& 	0.8925	&		$-$		\\  

\hline 
		1230.0042	& 		0.9144	&		$-$		\\

\hline 
		761.9995	&$-$	 	 	&	0.2654	\\   
\hline 
		1006.0133	& $-$		 	&	0.3176	\\   
\hline 
			1607.0092& 	$-$	 	&0.3378		\\   
\hline
\end{tabular}
\end{center}
\caption{Valores  experimentales  para los $T_{M}$ y $T_{m }$  del espectro de la Figura \ref{F4}.}
\end{table}

\begin{figure}[ht]
	\centering
			\includegraphics[width=0.5\textwidth]{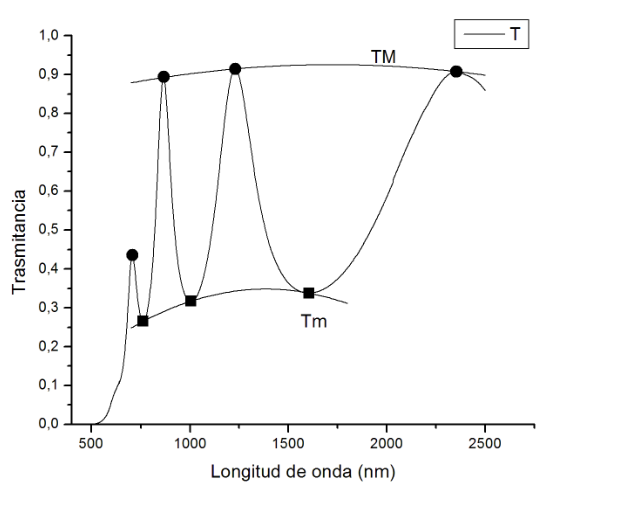}
		\caption{Se aprecia el espectro de transmitancia para la pel{\'i}cula de Silicio compensada, en donde se trazan los puntos $T_{M}$ y $T_{m}$ experimentales, junto con las  curvas envolventes que se  determinaron por el m{\'e}todo de interpolaci{\'o}n de Lagrange.}
	\label{F9}
\end{figure}

De acuerdo con los  valores  consignados  en el Cuadro I para el espectro de transmitancia de la Figura \ref{F4}, se calcularon los valores intermedios  de $T_{M}$ y de $T_{m}$ por el m{\'e}todo de interpolaci{\'o}n de Lagrange, ver Ap{\'e}ndice A. Este m{\'e}todo permite calcular las curvas envolventes del espectro en cuesti{\'o}n, tal como se  aprecia en la  Figura \ref{F9}.

Se aprecia que la regi{\'o}n de baja y media absorci{\'o}n corresponde al intervalo de
\begin{equation}
	700 nm\leq \lambda \leq 1800 nm.
	\label{Eq44}
\end{equation}
Para la regi{\'o}n de baja y media absorci{\'o}n establecida es posible  calcular  el {\'i}ndice  de refracci{\'o}n de la pel{\'i}cula $n(\lambda)$ por unidad de longitud de onda a partir de (34). Esto permiti{\'o}  construir la gr{\'a}fica de $n(\lambda)$ vs $\lambda$ que se observa en la  Figura ref{F9}.

\begin{figure}[ht]
	\centering
			\includegraphics[width=0.5\textwidth]{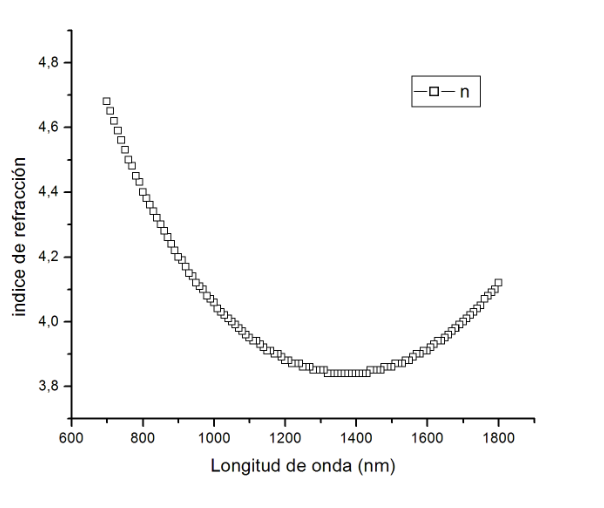}
		\caption{Comportamiento de {\'i}ndice de refracci{\'o}n $n(\lambda)$ en funci{\'o}n de  la longitud de onda $\lambda$ para el intervalo $700 nm\leq \lambda \leq 1800 nm$.}
	\label{F10}
\end{figure}
\begin{figure}[ht]
	\centering
			\includegraphics[width=0.5\textwidth]{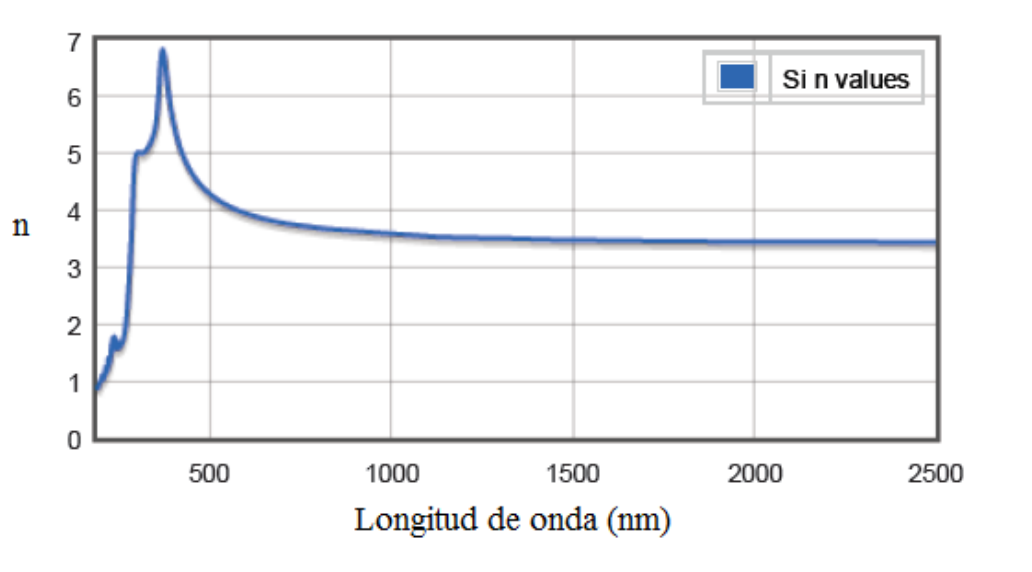}
		\caption{Comportamiento del {\'i}ndice refracci{\'o}n $n(\lambda)$ vs $\lambda$ para el Silicio, Tomado de Handbook
of Optical Constants of Solids, Edward D. Palik. Academic Press, Boston, 1985.}
	\label{F11}
\end{figure}
\begin{figure}[ht]
	\centering
			\includegraphics[width=0.5\textwidth]{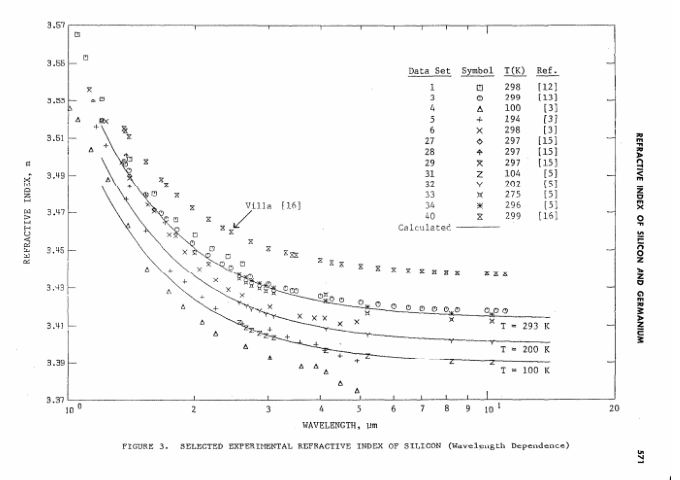}
		\caption{Comportamiento del {\'i}ndice refracci{\'o}n $n(\lambda)$ vs $\lambda$ para el Silicio, Tomado de \url{http://www.nist.gov/data/PDFfiles/jpcrd162.pdf}}
	\label{F12}
\end{figure}
\begin{figure}[ht]
	\centering
			\includegraphics[width=0.5\textwidth]{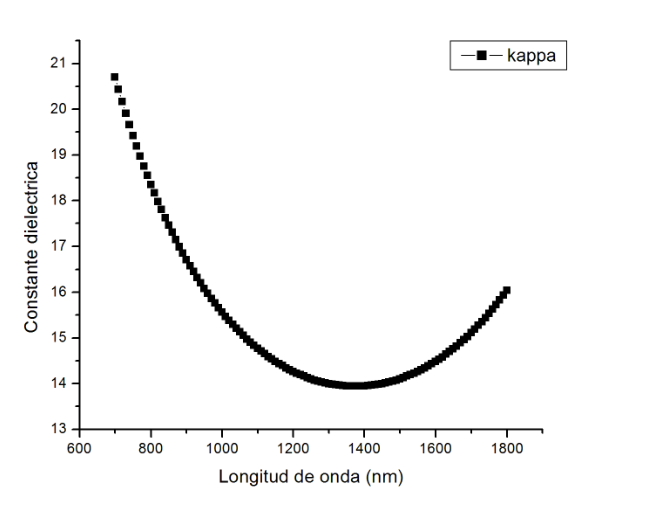}
		\caption{Comportamiento de la  constante diel{\'e}ctrica $\kappa$ en funci{\'o}n de  la longitud de onda $\lambda$ para el intervalo $700 nm\leq \lambda \leq 1800 nm$.}
	\label{F13}
\end{figure}
\begin{figure}[ht]
	\centering
			\includegraphics[width=0.5\textwidth]{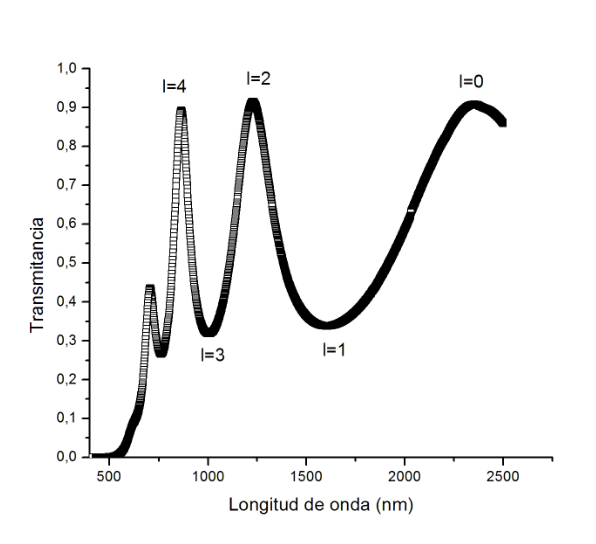}
		\caption{Puntos m{\'a}ximos  de orden $l$ .}
	\label{F15}
\end{figure}
\begin{figure}[ht]
	\centering
			\includegraphics[width=0.5\textwidth]{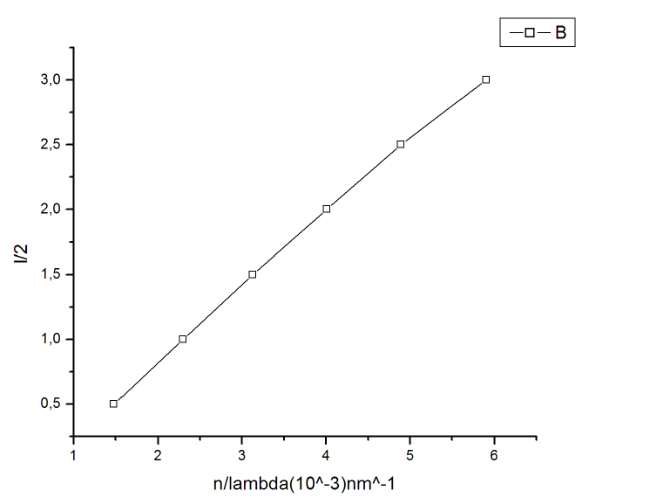}
		\caption{Gr{\'a}fica de $l/2$ vs $n/\lambda$ para determinar el numero de orden y el espesor la pel{\'i}cula.}
	\label{F16}
\end{figure}

Se puede realizar un comparativo del comportamiento del {\'i}ndice de refracci{\'o}n $n(\lambda)$ de la  pel{\'i}cula en estudio y el comportamiento del indice refracci{\'o}n para el Silicio puro reportado en la literatura \cite{Palik}, el  cual se puede ver en la Figura \ref{F11} y \ref{F12}. De la comparaci{\'o}n entre las  Figuras \ref{F10}, \ref{F11} y \ref{F12} se observa un comportamiento creciente del {\'i}ndice de refracci{\'o}n para bajas longitudes de onda en ambas gr{\'a}ficas. La  diferencia  radica en  el tipo de concentraci{\'o}n y los elementos que se le han añadido a la  pel{\'i}cula en estudio. Por definici{\'o}n se conoce que
\begin{equation}
	nv=c
	\label{Eq45},
\end{equation}
\begin{equation}
	n \frac{1}{\sqrt{\epsilon \mu}}=\frac{1}{\sqrt{\epsilon_{0}\mu_{0}}},
	\label{Eq46}
\end{equation}
donde se tiene que $\epsilon_{i}$ y $\mu_{i}$ corresponden a la permitividad  y la permeabilidad del medio $i$ respectivamente. Si la pel{\'i}cula delgada de Silicio compensado no es magn{\'e}tica $\mu\approx \mu_{0}$. Por lo que {\'i}ndice de refracci{\'o}n de la pel{\'i}cula delgada se reduce a
\begin{equation}
n=\sqrt{\frac{\epsilon}{\epsilon_{0}}}.
	\label{Eq47}
\end{equation}
De lo anterior, se puede deducir  el comportamiento de la permitividad de la  pel{\'i}cula  delgada. Adem{\'a}s, es posible determinar la  constante diel{\'e}ctrica $\kappa$ para la pel{\'i}cula delgada est{\'a} dada por
\begin{equation}
	\kappa=\frac{\epsilon}{\epsilon_{0}}=n^{2},
	\label{Eq48}
\end{equation}
cuyo comportamiento se observa en la Figura \ref{F13}. En la literatura se halla  que el valor de la constante diel{\'e}ctrica $\kappa$ para el Silicio en estado puro corresponde a $\kappa=11.7$, y en la misma  Figura se muestra que para el intervalo de longitud de  onda tenemos que la constante diel{\'e}ctrica  est{\'a} en el intervalo de 
	\[13,94\ll \kappa \ll 20,7.
\]
Lo cual confirma que la constante diel{\'e}ctrica corresponde al mismo orden de magnitud con lo reportado en la literatura \cite{kappa}.
Se conoce  que  la  ecuaci{\'o}n de Cauchy es de la  forma
\begin{equation}
n=a\frac{1}{\lambda^{2}}+c,
	\label{Eq49}
\end{equation}
donde se tiene  que $a$ corresponde a la pendiente y $c$ es el punto de corte de la  recta con el eje $y$.
	\begin{figure}[ht]
		\centering
				\includegraphics[width=0.5\textwidth]{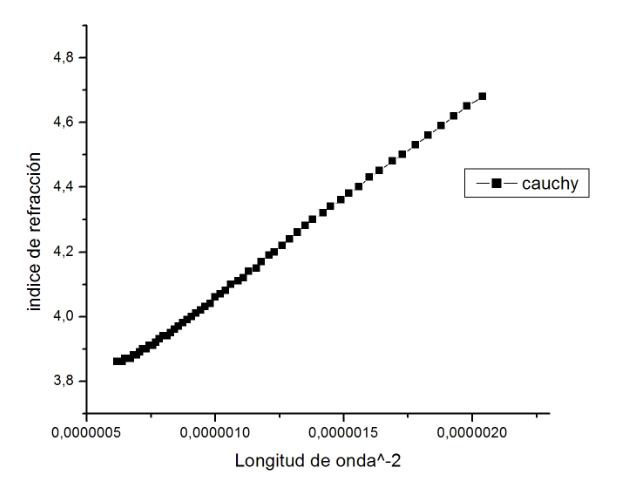}
			\caption{Comportamiento de {\'i}ndice de refracci{\'o}n $n(\lambda)$ en funci{\'o}n de   $\frac{1}{\lambda^{2}}$ para el intervalo $700 nm\leq \lambda \leq 1800 nm$ de  acuerdo a la  ecuaci{\'o}n de Cauchy.}
		\label{F14}
	\end{figure}
Por el m{\'e}todo de m{\'i}nimos cuadrados se determin{\'o} que la  ecuaci{\'o}n para los puntos hallados  con \eqref{Eq49} es de la  forma
\begin{equation}
n(\lambda)=\frac{596000}{\lambda^{2}}+3.4672,\,\,\,\,R=0.9993.
\label{Eq50}
\end{equation}
Posteriormente, con el espectro en estudio se determinaron los puntos m{\'a}ximos  de orden $l$ que se observan en la Figura \ref{F15}.
Esto permite hallar de una  manera indirecta el espesor  de la pel{\'i}cula  delgada de Silicio compensado. Dada la relaci{\'o}n
\begin{equation}
\frac{l}{2}=2d\left[\frac{n}{\lambda}\right]-m_{1}.
\label{Eq51}
\end{equation}
La  cual es representada en la Figura \ref{F16}. 
\begin{table}[!hbt]
\begin{center}
\begin{tabular}{|l | l | l|}
\hline
$\lambda(nm)$	 	& $n/\lambda\left(*10^{-3}\right)nm^{-1}$		&$l/2$ \\   
\hline 
		760	& 		5.9	&	$3.0$		\\   
\hline 
		870	& 	4.89	&		$2.5$		\\  
		
\hline 
		1010	& 		4.01	&		$2.0$		\\  

\hline 
		1230	& 		3.13	&		$1.5$		\\

\hline 
		1610	&$2.30$	 	 	&1.0	\\   
\hline 
		2350& $1.48$		 	&	0.5	\\   

\hline
\end{tabular}
\end{center}
\caption{Valores  experimentales  para los $n/\lambda$ y $ l$.}
\end{table}

Posteriormente, se representan los datos  consignados  en el Cuadro II que permite hallar de  manera indirecta el espesor  de la pel{\'i}cula $d$.
Donde se  calcul{\'o} el ajuste de la recta para  la Figura \ref{F16} que corresponde  a
\begin{equation}
\frac{l}{2}=2\left(282.98nm\right)\left[\frac{n}{\lambda}\right]-0.3
\label{Eq52}
\end{equation}
donde se identifican los t{\'e}rminos
\begin{equation}
m_{1}=0,3,\,\,\,\,\,2d=565.97nm.
\label{Eq53}
\end{equation}
De lo cual se puede afirmar  que el espesor te{\'o}rico  es $d=282.98nm$ y $m_{1}\approx 0,5$.

\begin{figure}[ht]
	\centering
			\includegraphics[width=0.5\textwidth]{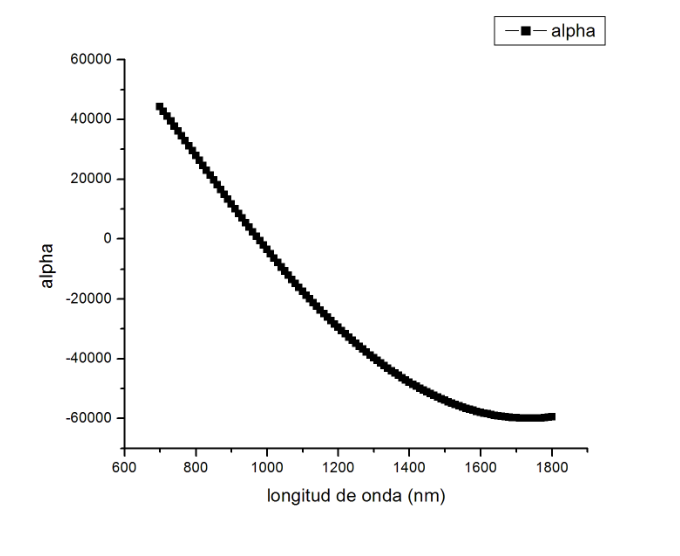}
		\caption{Comportamiento del coeficiente de absorci{\'o}n $\alpha$ en funci{\'o}n de la  longitud de onda $\lambda$.}
	\label{F17}
\end{figure}

\begin{figure}[ht]
	\centering
			\includegraphics[width=0.5\textwidth]{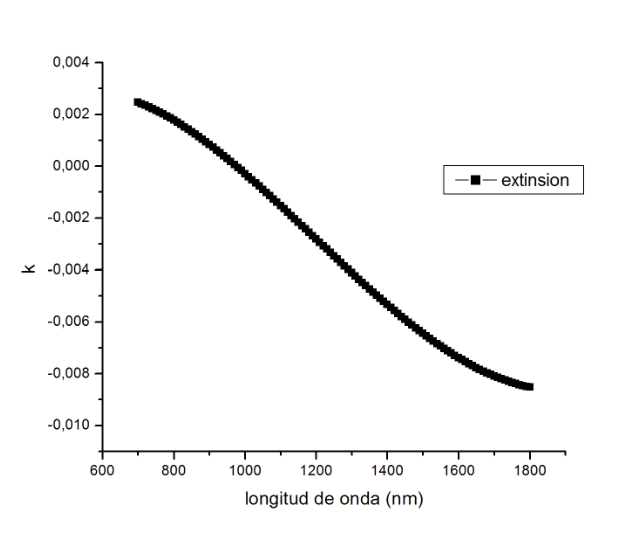}
		\caption{Comportamiento del coeficiente de  extinci{\'o}n $k$ en funci{\'o}n de la  longitud de onda $\lambda$.}
	\label{F18}
\end{figure}

\begin{figure}[ht]
	\centering
			\includegraphics[width=0.5\textwidth]{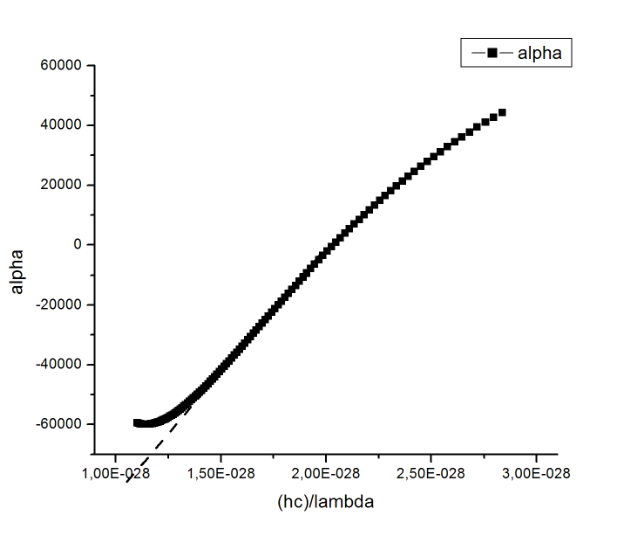}
		\caption{Determinaci{\'o}n de la banda de energ{\'i}a prohibida $E_{g}$ .}
	\label{F19}
\end{figure}

\begin{figure}[ht]
	\centering
			\includegraphics[width=0.5\textwidth]{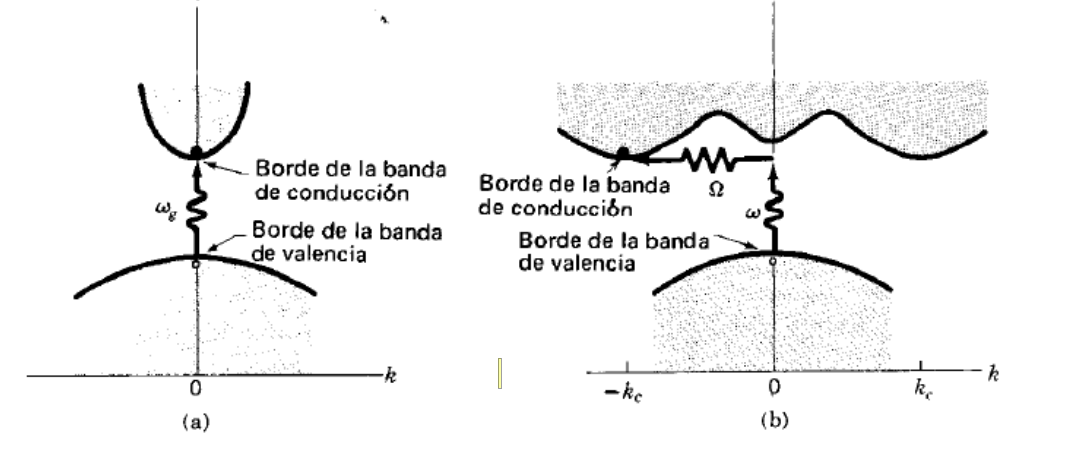}
		\caption{En (a) se presenta el proceso de transici{\'o}n entre la banda de valencia y la banda de conducci{\'o}n para un material de gap directo, tal transici{\'o}n se realiza por la interacci{\'o}n electr{\'o}n-fot{\'o}n. En (b) se involucra  tanto un fot{\'o}n como un fon{\'o}n para  que se de la transici{\'o}n de un electr{\'o}n de la banda de valencia a la banda de conduccion. Tomado de  \cite{Kittel}.}
	\label{F20}
\end{figure}

Posteriormente,  se calcula el coeficiente de absorci{\'o}n $\alpha(\lambda)$ a partir de \eqref{Eq22} y \eqref{Eq36}. En la Figura \ref{F17} se observa  el comportamiento del coeficiente de absorci{\'o}n $\alpha$ en funci{\'o}n de la  longitud de onda. Se tiene que el coeficiente de absorci{\'o}n aumenta con la disminuci{\'o}n de la longitud de onda. Esto puede deberse a que se est{\'a} aproximando  a la frecuencia natural de la pel{\'i}cula.
El siguiente paso, es calcular  el coeficiente de  extinci{\'o}n $k$ a  partir  de \eqref{Eq1}  y cuyo  comportamiento en funci{\'o}n de $\lambda$ se aprecia en la Figura \ref{F18}.
De  las  Figuras \ref{F11} y \ref{F18} es posible determinar el {\'i}ndice de refracci{\'o}n complejo $\eta=n-ik$ por unidad de longitud de onda.
Para la determinaci{\'o}n de la banda de energ{\'i}a prohibida $E_{g}$ se recurre a representar los valores del coeficiente de absorci{\'o}n $\alpha(\lambda)$ vs $E=\frac{hc}{\lambda}$ que se observan en la Figura \ref{F19}.
Lo cual permite realizar una  extrapolaci{\'o}n de la  regi{\'o}n que presenta un  comportamiento lineal y determinar el punto de corte de una recta dibujada sobre dicha regi{\'o}n y el eje $E=\frac{hc}{\lambda}$. Tal punto corresponde a
\begin{equation}
E_{g}=1.1655*10^{-28}J\approx 0.7275 eV.
\label{Eq54}
\end{equation}
De acuerdo a Kittel  el valor reportado del gap de Silicio puro corresponde  a $1.11 eV$  a una temperatura de $300K$ \cite{Kittel}. La diferencia entre el valor reportado y el hallado en el presente informe se deben atribuir  a dos posibles razones:
\begin{itemize}
	\item Margen de error en el presente calcul{\'o} dado que el error para el gap de la muestra en estudio  corresponde a $2.842\%$ que se puede ver en el Ap{\'e}ndice B.
	\item Nivel de concentraci{\'o}n de las impurezas u otros elementos presentes en la muestra en estudio.	
\end{itemize}
La presencia de impurezas permiten que se presenten transiciones de los portadores de carga de menor energ{\'i}a entre la banda de valencia y la banda de conducci{\'o}n. Para el presente estudio no es posible determinar qu{\'e} tipo de portadores de carga hay presentes en la  muestra, ello se podr{\'a} determinar en otro tipo de pruebas  tales  como someter la pel{\'i}cula  a un estudio de efecto Hall.  Se sabe que en un semiconductor se denomina de gap directo cuando el m{\'i}nimo de la banda de conducci{\'o}n y el m{\'a}ximo de la banda de valencia se dan para el mismo valor de $k$. De lo contrario es de gap indirecto, como  se sabe el silicio y el germanio poseen un gap indirecto \cite{Kittel}. En otras palabras el gap directo se caracteriza por la transferencia de electrones de la banda de valencia a la banda de conducci{\'o}n mediante la interacci{\'o}n solo de electr{\'o}n-fot{\'o}n. En el caso del gap indirecto ademas de la interacci{\'o}n electr{\'o}n-fot{\'o}n la transici{\'o}n de a  la  banda de conducci{\'o}n debe estar mediada por un fon{\'o}n \cite{Kittel}. Tal como se observa en la Figura \ref{F20}.
La absorci{\'o}n directa de un fot{\'o}n es una transici{\'o}n vertical desde la banda de valencia a la de conducci{\'o}n. Cuando un electr{\'o}n hace una transici{\'o}n vertical desde la banda de conducci{\'o}n a la de valencia emite un fot{\'o}n.  Tambi{\'e}n son posibles las transiciones indirectas que consisten en la absorci{\'o}n/emisi{\'o}n simult{\'a}nea de un fot{\'o}n y un fon{\'o}n. El salto vertical corresponde al fot{\'o}n y el cambio de $k$ se
debe al fon{\'o}n. La transici{\'o}n indirecta tiene una frecuencia mucho menor que la directa\cite{gap}. 
\section{Conclusiones}
Del presente  estudio realizado a una pel{\'i}cula delgada de Silicio compensado se extraen las siguientes  conclusiones:
\begin{enumerate}
	\item Se estudi{\'o} la pel{\'i}cula suministrada,  a  la  cual se le determinaron las  constantes  {\'o}pticas bajo el modelo propuesto por Swanepoel. La determinaci{\'o}n de los $T_{M}$ y $T_{m}$ se realiz{\'o} por interpolaci{\'o}n de Lagrange. Esto permiti{\'o} determinar una  ecuaci{\'o}n de  Cauchy de la  forma
	\begin{equation}
		n(\lambda)=\frac{596000}{\lambda^{2}}+3.4672.
		\label{Eq55}
	\end{equation}
	\item A la pel{\'i}cula en estudio se le practicaron 5 ensayos de perfilom{\'e}tria en los que se  determin{\'o} que el espesor es
			\begin{equation}
			d_{p}=308.4998nm.
	\label{Eq56}
	\end{equation}
	Por otro lado empleando el modelo propuesto, de acuerdo  a la Figura \ref{F13} y por \ref{F14}, se determin{\'o} que el valor  $d$  es de 
	\begin{equation}
d=282,96nm\pm 3\%.
	\label{Eq57}
	\end{equation}
	Y el error es del $8\%$.
	\item El indice refracci{\'o}n $s$ para el sustrato que en este caso corresponde al vidrio est{\'a} dentro del orden de magnitud reportado en la literatura 
		\begin{equation}
	s=1,5142.
	\label{Eq58}
	\end{equation}
	\item De la comparaci{\'o}n entre las  Figuras \ref{F10}, \ref{F11} y \ref{F13} se observa un comportamiento creciente del {\'i}ndice de refracci{\'o}n para bajas longitudes de onda en ambas gr{\'a}ficas. La  diferencia  radica en la el tipo de concentraci{\'o}n y el elemento que se le ha añadido a la  pel{\'i}cula en estudio.
	\item Se estudio el comportamiento de la constante diel{\'e}ctrica, su relaci{\'o}n con el indice de refracci{\'o}n. Hall{\'a}ndose que el valor de la constante diel{\'e}ctrica para una pel{\'i}cula delgada de Silicio compensado esta de acuerdo con lo reportado en la literatura.
	\item El valor gap reportado para el Silicio en puro corresponde a $1,11eV$ y el valor del gap para la muestra estudiada corresponde a $0.7275eV$, la discrepancia se debe al nivel del concentraci{\'o}n con que esta  compensado la pel{\'i}cula de Silicio.
	\end{enumerate}
\appendix
\section{Polinomios de interporlaci{\'o}n de Lagrange}
Los polinomios de interpolaci{\'o}n de Lagrange corresponden a uno de los diversos m{\'e}todos que  para $n+1$ puntos en un plano con diferentes coordenadas de x, permiten encontrar una funci{\'o}n polin{\'o}mica del menor grado posible que pase por esos $n+1$ puntos. Tales polinomios son escritos de la forma
\begin{equation}
P(x)=\sum_{i=0}^n l_{i}(x)y_{i},
\end{equation}
donde
\begin{equation}
l_{i}(x)=\prod_{m\neq i,m=0}^{n} \dfrac{x-x_{m}}{x_{i}-x_{m}}.
\end{equation}
Y al expandir la productoria se tiene
	\begin{widetext}
\begin{equation}
l_{i}(x)=\prod_{m\neq i,m=0}^{n} \dfrac{x-x_{m}}{x_{i}-x_{m}}=\dfrac{x-x_{0}}{x_{i}-x_{0}}\cdots\dfrac{x-x_{i-1}}{x_{i}-x_{i-1}}\cdot\dfrac{x-x_{i+1}}{x_{i}-x_{i+1}}\cdots \dfrac{x-x_{n}}{x_{i}-x_{n}}
\end{equation}
\end{widetext}
Estos polinomios son construidos con una propiedad importante que dice,
\begin{equation}
l_{i}(x_{j})=\delta_{ij}.
\end{equation}
Este m{\'e}todo, ofrece un mejor ajuste de los puntos $T_{M}$ y $T_{m}$ que fueron hallados experimentalmente para el espectro de transmitancia de la Figura \ref{F4}.

Tales polinomios de interpolaci{\'o}n puesto que para el espectro obtenido de la medici{\'o}n de la muestra de silicio compensado en partes por mill{\'o}n, se quiere encontrar una funci{\'o}n de la envolvente del espectro con base en los m{\'a}ximos y m{\'i}nimos de transmitancia obtenidos de la medici{\'o}n que en este caso es de $3$ m{\'a}ximos y $3$ m{\'i}nimos. 
Para el presente trabajo realizado los polinomios de interpolaci{\'o}n de Lagrange se interpretan de la siguiente manera:
\begin{enumerate}
	\item La variable $x$ corresponde a la longitud de onda del espectro obtenido en la medici{\'o}n.
	\item $x$ corresponde a la longitud de onda del espectro obtenido en la medici{\'o}n,
	\item  $x_m$ corresponde a la longitud de onda siguiente o anterior a la longitud de onda correspondiente a cada m{\'a}ximos o m{\'i}nimos de transmitancia medidos en el espectro de la muestra del punto que se est{\'a} utilizando para la interpolaci{\'o}n,
	\item $x_i$ corresponde a la longitud de onda del punto que se est{\'a} interpolando y
	\item $y_i$ es el valor de la transmitancia que corresponde al punto xi que se est{\'a} analizando.
\end{enumerate}

\section{An{\'a}lisis de incertidumbre y error}

\subsection{Incertidumbre en las mediciones}

Para realizar el an{\'a}lisis en la incertidumbre en las mediciones se consideran los siguientes casos:

\begin{itemize}
 \item [$1.$] Si la propagaci{\'o}n del error proviene de una ecuaci{\'o}n se aplica la relaci{\'o}n

\begin{equation}
\Delta f=\sum_{i=1}^{n}\left\arrowvert\frac{\partial f}{\partial x_i}\Delta x_i\right\arrowvert .
\end{equation}

Siendo $\Delta x_i$ la incertidumbre en las mediciones y $\Delta f$ la incertidumbre propagada. Esta es la forma en que la que se hallar{\'a} el error para $n$, $\alpha$ y $k$.
 \end{itemize}

\begin{itemize}
 \item [$2.$] Si la medida proviene de promediar un conjunto de datos, la incertidumbre es la desviaci{\'o}n est{\'a}ndar, calculada f{\'a}cilmente a trav{\'e}s de herramientas de software. 

\end{itemize}

\begin{itemize}
 \item [$3.$] S{\'i} la medida se obtiene a partir de un m{\'e}todo de ajuste a una recta, como es el caso del espesor de la pel{\'i}cula, $d$, y la energ{\'i}a de gap $E_g$. Se aplica la t{\'e}cnica que se describe a continuaci{\'o}n:
 Se hace un promedio de los errores de las variables ubicadas en los ejes $x$ e $y$ en la gr{\'a}fica
\begin{equation}
 \sigma_x=\frac{\sum_{i=1}^{n} x_i}{n} 
 \end{equation}
 \begin{equation}
  \sigma_y=\frac{\sum_{i=1}^{n} y_i}{n}
 \end{equation}
A partir de los cuales se halla un par{\'a}metro $\sigma$
\begin{equation}
 \sigma=\sqrt{\sigma_y^2+m^2\sigma_x^2}
\end{equation}
Con $m$ la pendiente de la recta. A partir de este par{\'a}metro se hallan la incertidumbre en la pendiente y en el intercepto con el \textit{eje y}.
\begin{equation}
\Delta m=\sqrt{\frac{n\sigma^2}{n \sum_{i=1}^{n} x_i^2-\left(\sum_{i=1}^{n} x_i \right)^2 }} 
\end{equation}
\begin{equation}
\Delta c=\sqrt{\frac{\sigma^2 \sum_{i=1}^{n} x_i^2}{n \sum_{i=1}^{n} x_i^2-\left(\sum_{i=1}^{n} x_i \right)^2 }}  
\end{equation}
\end{itemize}
Aplicando los m{\'e}todos descritos seg{\'u}n los casos correspondientes, se halla primero una incertidumbre para $n$ a partir de 
\begin{equation}
n(\lambda)=\frac{596000}{\lambda^{2}}+3,3672.
\end{equation}
bajo la hip{\'o}tesis que se hall{\'o} una relaci{\'o}n anal{\'i}tica entre $n$ y $\lambda$ dada por  \eqref{Eq50}

\begin{equation}
\Delta n=\frac{1.19*10^6 \Delta\lambda}{\lambda^3} 
\end{equation}
Seg{\'u}n especificaciones t{\'e}cnicas $\Delta\lambda=0,02 nm$ para el espectrof{\'o}metro Cary 5000, lo cual permite a su vez hallar la incertidumbre de $\frac{n}{\lambda}$
\begin{equation}
\Delta \left(\frac{n}{\lambda} \right)=\frac{\Delta n}{\lambda}+\frac{n \Delta \lambda}{n^2}
\end{equation}
Esto corresponde al error en los datos del \textit{eje x} para el m{\'e}todo gr{\'a}fico que permiten determinar el espesor de la pel{\'i}cula $d_{Teo}$. Posteriormente se aplica el m{\'e}todo 3, teniendo en cuenta que $\sigma_y=0$, pues en el \textit{eje y}  s{\'o}lo se ubican valores $\frac{l}{2}$ que no llevan ning{\'u}n error inherente, se encuentra la incertidumbre en la pendiente. Dado que
\begin{equation}
\Delta d=\frac{\Delta m}{2}.
\end{equation}
As{\'i} se encuentra que la incertidumbre para el espesor es de 
\begin{equation}
\Delta d=8,447 \ nm,\,\,\,\, \Delta d=2,985\%.
\end{equation}
Los datos calculados  a partir de (37) para $x$ se observ{\'o} que todos son cercanos $1$. Por lo cual, es posible aplicar el m{\'e}todo 2 y considerar el error en $x$ como su desviaci{\'o}n est{\'a}ndar. Se halla que
\begin{equation}
\Delta x=0,009,\,\,\,\,\Delta x=0,919\%. 
\end{equation}
Utilizando el m{\'e}todo 1 se pueden encontrar las incertidumbres en $\alpha$ y $k$
\begin{equation}
 \Delta \alpha=\frac{lnx \Delta d}{d^2}+\frac{\Delta x}{xd}
\end{equation}
\begin{equation}
 \Delta k=\frac{1}{4\pi}(\Delta \alpha \lambda+\alpha \Delta \lambda)
\end{equation}
Dado que el c{\'a}lculo en el gap proviene de un m{\'e}todo gr{\'a}fico, en el cual el valor de esta energ{\'i}a es el intercepto en el \textit{eje x}, se utiliza el m{\'e}todo 3, teniendo en cuenta que, para poder aplicar correctamente 
\begin{equation}
\Delta c=\sqrt{\frac{\sigma^2 \sum_{i=1}^{n} x_i^2}{n \sum_{i=1}^{n} x_i^2-\left(\sum_{i=1}^{n} x_i \right)^2 }},  
\end{equation}
donde $\Delta \alpha$ son los errores $\Delta x_i$y $\Delta\left(\frac{hc}{\lambda}\right)$ son los errores $\Delta y_i$. Calculando los $\Delta x_i$ y $\Delta\left(\frac{hc}{\lambda}\right)$ a partir de 
\begin{equation}
\Delta \left( \frac{hc}{\lambda}\right)=\frac{hc\Delta \lambda}{\lambda^2}.
\end{equation}
Por otro lado, se tiene que la incertidumbre en el gap corresponde a 
\begin{equation}
\Delta E_g=0,0357eV ,\,\,\,\, \Delta E_g=2,842 \%.
\end{equation}
\subsection{C{\'a}lculo de error}
El error $\epsilon$ en la medici{\'o}n se determina mediante
\begin{equation}
\epsilon=\left|\frac{V_{m}-V_{r}}{V_{r}}\right|,
\end{equation}
donde $V_{m}$ corresponde al valor obtenido en la medici{\'o}n y $V_{r}$ es un valor que se usa como referencia.  Para el valor de referencia del espesor de la pel{\'i}cula se realiz{\'o} un conjunto de $5$ mediciones de \textit{perfilometr{\'i}a}. Promediando los datos de las $5$ mediciones a partir de los datos entre la zona de la medici{\'o}n entre los $4000 \AA$ y los $5000 \AA$  se obtiene un espesor de la pel{\'i}cula de Silicio compensado
\begin{equation}
d_{r}=308.500 nm. 
\end{equation}
A partir del cual el error en la medici{\'o}n del espesor es de
\begin{equation}
 d=25.519 \ nm,\,\,\,\,\,\, \epsilon= 8.272 \%
\end{equation}
Teniendo en cuenta la incertidumbre en la medida del espesor, el error puede reducirse hasta $5.287\%$.
\section*{References}
\providecommand{\newblock}{}


\begin{thebibliography}{10}
\expandafter\ifx\csname url\endcsname\relax
  \def\url#1{{\tt #1}}\fi
\expandafter\ifx\csname urlprefix\endcsname\relax\def\urlprefix{URL }\fi
\providecommand{\eprint}[2][]{\url{#2}}
\bibitem{Swanepoel}  	R Swanepoel.  J. Phys. E: Sci. Instrum. 16 (1983) 1214
\bibitem{Mesa} Mesa F, Ballesteros V, Dussan A (2014) C{\'a}lculo de constantes {\'o}pticas de pel{\'i}culas delgadas de Cu3BiS3 a trav{\'e}s
del m{\'e}todo de Wolfe. Universitas Scientiarum 19(2): 123-131
doi: 10.11144/Javeriana.SC19-2.ccop
\bibitem{Kittel} Kittel, C., Introduction to Solid State Physics, 6th Ed., New York:John Wiley, 1986.
\bibitem{Garzon}D. Garzon , A. Dussan , J. Malambo. Revista Colombiana de F{\'i}sica, Vol. 43, No. 2 de 2011. 	
\bibitem{Lagrange} Polinomios de interpolacion de Lagrange. Consultado mayo del 2015, disponible en: \url{http://www.matematicasvisuales.com/html/analisis/interpolacion/lagrange.html}.
\bibitem{Palik}  E D. Palik. Handbook of Optical Constants of Solids, . Academic Press, Boston, 1985
\bibitem{indice} Indice de refracci{\'o}n del Silicio. Consultado mayo del 2015, disponible en: \url{http://www.nist.gov/data/PDFfiles/jpcrd162.pdf}
\bibitem{kappa} Constante diel{\'e}ctrica para el Silicio. Consultado mayo del 2015, disponible en: \url{http://www.ioffe.ru/SVA/NSM/Semicond/Si/basic.html}
\bibitem{Akai} T. J. Akai. M{\'e}todos  Num{\'e}ricos aplicados a la ingenier{\'i}a. Limusa Wiley, 2000.
\bibitem{udea}Universidad de Antioquia. Introducci{\'o}n al m{\'e}todo de m{\'i}nimos cuadrados. Medell{\'i}n, 2010.
\bibitem{gap} Gap directo-indirecto. Consultado mayo del 2015, disponible en: \url{http://ocw.upc.edu/sites/default/files/materials/15014928/3.-_semiconductores-4826.pdf}
\end{thebibliography}
\end{document}